# Unconventional superconductivity induced in Nb films by adsorbed chiral molecules

H. Alpern[1,2], E. Katzir[2], S. Yochelis[2], N. Katz[1], Y. Paltiel*[2], and O. Millo*[1]

[1]*Racah Institute of Physics and the Center for Nanoscience and Nanothechnology, The Hebrew University of Jerusalem, Jerusalem 91904, Israel*

[2]*Applied Physics Department and the Center for Nanoscience and Nanothechnology, The Hebrew University of Jerusalem, Jerusalem 91904, Israel*

**Abstract**

Motivated by recent observations of chiral-induced magnetization and spin-selective transport we studied the effect of chiral molecules on conventional BCS superconductors. By applying scanning tunneling spectroscopy, we demonstrate that the singlet-pairing s-wave order parameter of Nb is significantly altered upon adsorption of chiral polyalanine alpha-helix molecules on its surface. The tunneling spectra exhibit zero-bias conductance peaks embedded inside gaps or gap-like features, suggesting the emergence of unconventional triplet-pairing components with either d-wave or p-wave symmetry, as corroborated by simulations. These results may open a way for realizing simple superconducting spintroinics devices.

## 1. Introduction

There are rather few materials showing signature of spin-triplet pairing superconductivity. One notable example is $Sr_2RuO_4$, for which the superconducting order parameter (OP) was suggested to have chiral p-wave symmetry[1,2]. Mixed spin-singlet and spin-triplet pairing was observed in non-centrosymmetric heavy fermion superconductors, like $CePt_3Si$ [3]. An important ingredient in the theories addressing the latter phenomenon (which may be relevant also to our results) is the required contribution of Rashba type anti-symmetric spin-orbit coupling [4,5]. In addition to fundamental interest, triplet superconductors are attractive also as components in superconducting spintronic devices [6]. However, the transition temperature, $T_C$, in these systems is very low (<1.5 K; the $T_C$ of $Sr_2RuO_4$), and high quality single crystals are typically required.

Triplet superconductivity at much higher temperatures can be achieved by coupling a spin-singlet superconductor to a ferromagnet. Such hybrid superconductor-ferromagnetic systems have been studied thoroughly [6], but mainly in the context of spin-triplet superconducting correlations induced in the ferromagnet. The inverse problem, in which a triplet-pairing component is proximity-induced in a pristine singlet superconductor, was addressed in some theoretical works, predicting various pairing-potential symmetries: odd-frequency s-wave [7,8] or a combination of odd-frequency s-wave and d-wave with even-frequency p-wave [9], all respecting fermionic anti-symmetrisation. This approach has been proven successful in various experiments performed on hybrid superconductor/ferromagnet multilayers [10–12]. Typically, this required the application of sophisticated epitaxial growth techniques, and thus simpler fabrication methods should be sought for. Following the experimental evidence listed in the next paragraph, we turned to examine the possibility of inducing triplet-pairing superconductivity in a conventional singlet superconductor via the adsorption of chiral molecules.

Compelling evidence has been gained in recent years that electron transfer through chiral molecules is spin selective. Several experimental studies have clearly demonstrated that chiral molecules can serve as very efficient spin filters or polarizers, achieving spin selectivity of more than 60% [13,14]. Moreover, recent works demonstrated the ability of chiral molecules to effectively magnetize Ni layers attached to them[15]. Finally, we have also shown that the adsorption of organic molecules on the surface of Nb films using a simple wet chemistry procedure does not degrade their superconducting properties. This process may even increase the films' $T_C$ in cases where Au or Co nanoparticles are attached to the molecules [16,17].

In this work, we studied via STS the modification of the OP symmetry of a conventional singlet-pairing s-wave superconductor (Nb), upon the adsorption of chiral molecules (polyalanine alpha-helix) on its surface. The tunneling spectra measured on areas covered with molecules reflect unconventional superconductivity and can be accounted for by a combination of s-wave and either d-wave or triplet-pairing p-wave OP symmetries.

## 2. Experimental section

Nb films, 60 nm thick, with critical temperature ($T_C$) varied between 7.4 k and 8.5 k were sputtered on Si wafers. Some of the films were treated by Ar plasma asher in order to enhance the adsorption. The samples were then inserted into a solution of polyalanine alpha-helix molecules dissolved in ethanol for 24 h. More details regarding the sample preparation are presented in Supplementary Material.

The STM/STS measurements were conducted employing a home-built cryogenic STM that operates in a clean He exchange gas environment, using Pt-Ir tips. Topographic images were taken with bias voltages of 100 mV, well above the superconducting gap voltage of Nb. The tunneling conductance *dI/dV* versus *V* spectra (proportional to the local density of states, DoS), were acquired while momentarily disconnecting the STM feedback loop. The set current and bias-voltage values (before disabling the feedback) were typically 0.1 nA and 10 mV.

## 3. Results and discussion

Figure 1(a) portrays an STM topography image of an area covered by molecules, showing features of length varying between 1.5 and 3 nm. The topographic cross-section presented in figure 1(b), taken normal to a stack of parallel-aligned molecules (blue line in figure 1(a)), exhibits surface corrugation with amplitude of up to 1 nm and about 2 nm separation between peaks. These topographic features are consistent with the diameter (~1 nm) and length (~3 nm) of the polyalanine alpha-helix molecules we used [18,19], assuming they are not vertically attached to the surface, but rather tilted. In areas that exhibited such topographic features, we witnessed signs of unconventional superconductivity in the DoS. The adsorption process was not fully uniform, as evidenced by the image presented in figure 1(c) and the corresponding cross-section (figure 1(d)), measured on the same sample, showing two smooth planes separated by an atomic step, with no evidence of molecules. In such regions, we found only BCS-like gaps, free of in-gap features.

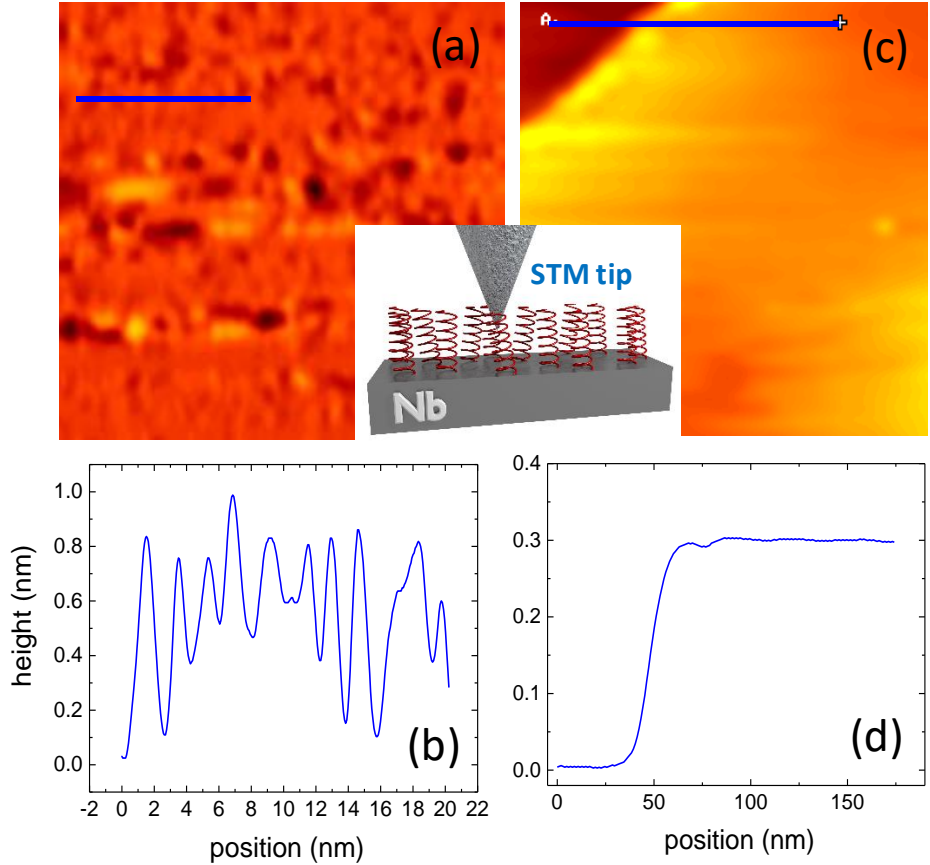

*Figure 1: Topographic images of a Nb film after polyalanine alpha-helix molecules adsorption. (a,b) Topographic image of a region covered with molecules and a cross-section taken along the blue line is presented in (b). In these regions the spectra reflected unconventional superconductivity. (c,d) Topographic image of a molecule-free region (c) along with a cross-section which exhibits an atomic step (d) taken along the blue line (c). In such areas we found only conventional BCS-like gaps. Inset: Measurement configuration scheme in a molecule covered region.*

In figure 2 we display the three typical types of *dI/dV-V* tunneling spectra (all measured at 4.2 K) found on each one of the eight samples we measured, although with different abundance. Figure 2(a) exhibits conventional BCS spectra with smeared features, found over the areas that did not show any traces of molecules. These spectra present the typical s-wave behavior expected for Nb, with no signs of modified OP symmetry. Figure 2(b) shows the first type of unconventional non-BCS spectra, which includes a small ZBCP situated inside a superconducting gap structure, signifying OP alteration. In some cases the coherence peaks are still visible – remnants of the original superconducting gap. Figure 2(c) displays the second type of non s-wave spectra, where a more distinctive ZBCP is embedded inside a broader gap-like feature that lacks visible coherence peaks. The two types of spectra presented in 2(b) and 2(c) were found macroscopically remote from one another, and retained the same features over distances of several dozen nanometers. All superconducting gaps and zero-bias anomalies in the DoS reduced with temperature and vanished above $T_C$, as shown in figure 3.

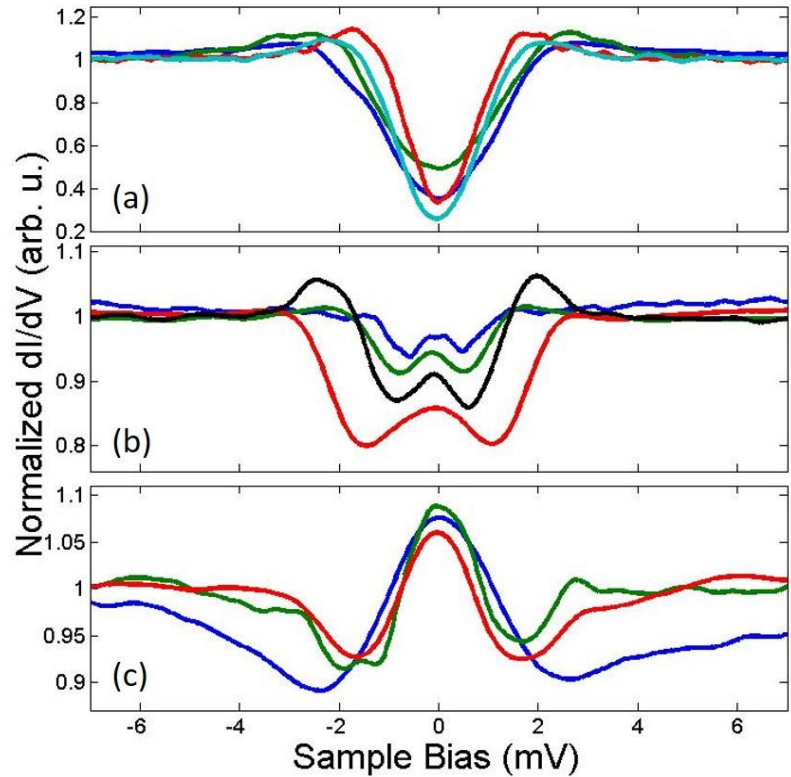

*Figure 2: A selection of three types of tunneling spectra measured over various regions and samples of Nb films after polyalanine alpha-helix molecules adsorption. (a) Conventional (but smeared) BCS s-wave spectra, exhibiting gaps with small coherence peaks measured on molecule-free regions. (b,c) Small ZBCPs inside a superconducting gap (b), and pronounced ZBCPs inside a gap-like feature (c), measured on regions with adsorbed molecules.*

The above spectral modifications were not sensitive to the specific pre-adsorption surface treatment or adsorption procedure applied to different samples, nor to the $T_C$. For example, the spectra presented in figure 2 were acquired from various samples, prepared with or without plasma ashing and from two different pristine Nb films having $T_C$ of 7.4 K to 8.5 K. Thus, the effect of the chiral molecules on the tunneling spectra, and thus the OP symmetry at the Nb surface, appears to be robust.

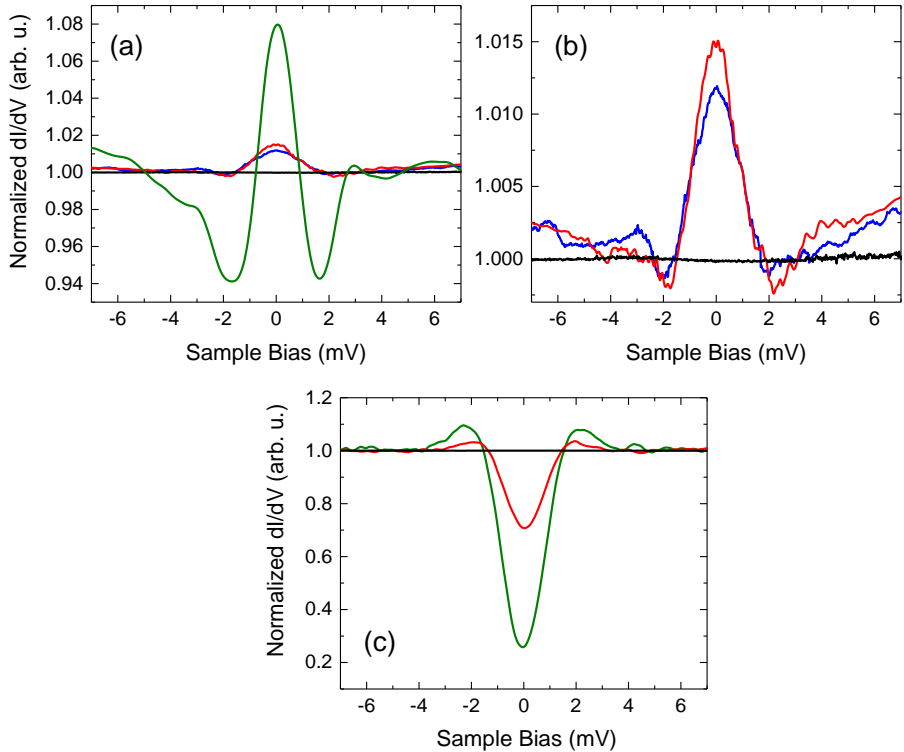

*Figure 3*: *Temperature dependence of the tunneling spectra. (a) Tunneling spectra measured at a molecule-covered area of a $T_C$ = 7.4 K sample at T = 4.2 K (green curve), T = 6.5 K (blue and red curves) and T = 15 K (black curve). (b) Blow up showing only the 6.5 K and 15 K data. (c) Tunneling spectra measured on the same sample, but at a molecule free region, at 4.2 K (green) and 6.5 K (red) and 15 K (black).*

We now turn to address the possible chiral-induced OP symmetries. This is done by fitting the measured tunneling spectra to the Blonder, Tinkham and Klapwijk (BTK) model [20], modified to treat other, non s-wave, OP symmetries as needed. We applied the BTK approach because in our case, due to the organic molecule adsorption, large tunneling barriers are not *a-priory* guarantied. Given a specific symmetry, the standard parameters of such fits are the pairing amplitude (or energy gap), $\Delta_0$, the tunnel barrier strength, Z, and the Dynes lifetime broadening parameter [21], $\Gamma$ (more details are given in the Supplementary Material) . All the theoretical spectra shown in figure 4 were calculated taking T = 4.2 K and Z ≥ 10. Such high Z values are well within the tunneling (rather than Andreev) spectroscopy regime in the BTK model, thus confirming that our spectra were acquired in *bona-fide* tunneling conditions. This is important to verify, since enhanced sub-gap conductance is observed also for s-wave superconductors in the low Z (typically < 1) regime, but such Andreev reflection spectra cannot account for the spectral features of narrow ZBCPs embedded in gaps shown in figures 2(b,c) ([20] and figure S1). Thus, we tried fitting our spectra using OP symmetries that are known to yield ZBCPs in the tunneling spectra, reflecting the formation of zero-energy Andreev bound states. Most notable are the anisotropic, sign-changing d-wave [22] and p-wave [23,24] OPs. It is well established that the existence and magnitude of ZBCPs in the tunneling spectra measured on d-wave or p-wave superconductors highly depend on the tunneling direction with respect to the crystallographic (and the corresponding pairing-potential) orientations. However, in case of proximity-induced unconventional superconductivity, the crystallographic orientation appears not to play any role [10,11,25–27]. This justifies the use of anisotropic pairing potentials in our case, albeit the isotropic nature of superconductivity in Nb and the disordered orientation of the adsorbed molecules.

Figure 4 presents three selected spectra (black solid lines), one of each type shown in figure 2, along with simulated spectra that gave the best fits. Figure 4(a) depicts a conventional gap structure measured on a molecule-free area, along with a corresponding fit calculated with merely s-wave OP. Evidently, the simulation fits the experimental data well, and a relatively small gap of $\Delta = 1.05$ $meV$ is extracted. This gap is smaller than that measured on high quality Nb samples, $\Delta \approx 1.55$ meV (Ref. [28] and figure 5(c)) and the coherence peaks are smeared, which is not surprising in view of the rather low $T_C$ of the pristine film (7.4 K). We note that molecules were not found within a few hundred nm from positions where such BCS-type spectra were measured, thus manifesting the localized nature of the chiral-induced OP modification.

Best fits to the spectra showing ZBCPs were obtained assuming a coexistence of s-wave and either d-wave ($d_{x^2-y^2}$) [22] or chiral p-wave ($p_x + ip_y$) [23] OP symmetries. Namely, the fits presented in figures 4(b,c) represent weighted sums of spectra calculated with the corresponding pairing-potentials. By using d-wave or p-wave pairing potentials only, we could fit the ZBCP features, but not the gap structures within which they are embedded; the s-wave contribution appears to be needed to achieve the latter task. For the $d_{x^2-y^2}$-wave simulations we used the standard form of the pairing amplitude, $\Delta$ = $\Delta_0\cos(2\alpha)$, where $\alpha$ is the angle between the tunneling and the pairing-potential lobe directions. Best fits involving p-wave OP was obtained with the chiral $(p_x + ip_y)$ pairing-potential $\Delta_{\uparrow\uparrow}(\theta,\phi) = \Delta_0\sin(\theta)e^{i\phi}$, while $\Delta_{\uparrow\downarrow}(\theta,\phi) = \Delta_{\downarrow\uparrow}(\theta,\phi) = \Delta_{\downarrow\downarrow}(\theta,\phi) = 0$. Here $\theta$ and $\phi$ are the polar and azimuthal angles, respectively. Importantly, this pairing potential[23,25] breaks time reversal symmetry, and is associated with a parallel-spin (m=1) triplet state. Simulations with m=0 pairing potential, $\Delta_{\uparrow\downarrow}(\theta,\phi), \Delta_{\downarrow\uparrow}(\theta,\phi) \neq 0$[24], and other p-wave related pairing potentials (e.g., $p_x$) proved not able to fit the measured data.

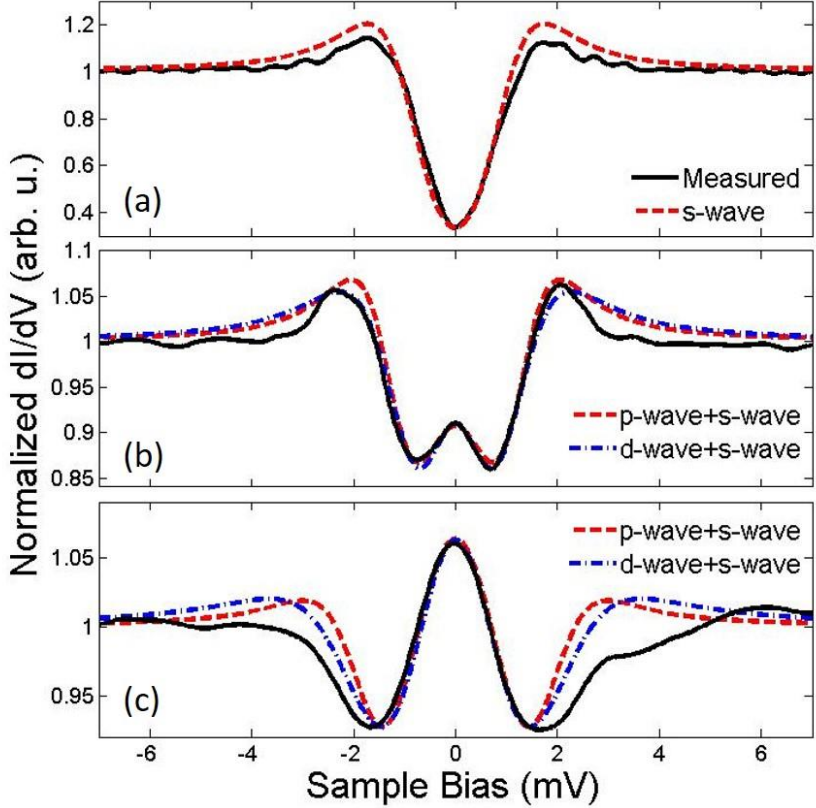

*Figure 4: The three different types of experimental tunneling spectra corresponding to figure 1 (black curves), and spectra calculated using the extended BTK model (dashed lines) with different combinations of order parameter symmetries. (a) s-wave, with Δ=1.05 meV and Γ = 0.08meV. (b) s-wave + p-wave (red): Δ=1.36 meV, s-wave weight 39% ; s-wave + d-wave (blue): Δ=1.3 meV, s-wave weight 50%. (c) s-wave + p-wave (red): Δ=1.9 meV, s-wave weight 27%; s-wave + d-wave (blue): Δ=2 meV, s-wave weight 35%.*

Figure 4(b) demonstrates that the second type of tunneling spectra can be very well accounted for by either one of the combined s+p or s+d OP symmetries described above. The corresponding calculated spectra fit equally well the ZBCP and gap structures of the experimental curve, with similar gap value, $\Delta_0$ = 1.36 meV (s+p) and 1.3 meV (s+d). The two fits differ notably, however, by the relative weight of the s-wave component, being 39% in the s+p simulated and spectrum (red dashed curve) and 50% in the s+d simulation (blue dot-dashed curve). This relative weight is an important parameter, as it provides insight into the effect of the chiral molecules in modifying the OP symmetry. The fits to the third type of tunneling spectra were not as good, as shown in figure 4(c). Both s+p and s+d fits grasp the ZBCP feature of the measured data quite accurately, but fail to match the relatively wide and asymmetrical measured gap-like features. As expected, the more pronounced ZBCP is expressed in a smaller weight of the s-wave component in the simulated spectra (compared to figure 3(b)); 27% and 35% in the s+p and s+d fits, respectively. We note that the anomalously wide gap-like features are observed nearly exclusively in spectra with pronounced ZBCPs, that is, when the non-conventional superconducting component is more significant. Possibly, in such cases the density of adsorbed chiral molecules is larger, thus interfering with the tunneling measurement.

We now turn to discuss other possible origins of the ZBCPs in our tunneling spectra. One may argue that the chiral polyaniline alpha-helix molecules induce upon adsorption an effective magnetic moment at the surface, giving rise to Yu-Shiba-Rusinov states [29–31] that are known to manifest themselves as in-gap peaks in the tunneling spectra. Although we cannot completely rule out this scenario, is appears less probable than our interpretation above. First, the molecules we use are free of any magnetic elements (see Supplementary Material). Second, the ZBCPs we observed *were always* practically symmetrically around zero bias, irrespective of position in the molecule-covered regions – on or between molecules. This is in vast contrast to theoretical predictions [29–31] and experimental observations [32,33]. Also Kondo-effect related resonances, as observed in Ref. [34] on a system comprising manganese-phthalocyanine (MnPc) molecules (or Mn) adsorbed on Pb(111), seems not to apply in our system due to the same reasons. We note in passing that the only different spectral feature (showing no ZBCP) we occasionally on molecular free regions, and on 4 samples only, was a very shallow and narrow gap lacking coherence peaks (figure S2); here too, showing no relation to a nearby magnetic impurity. This feature can be due to regions where the molecules are loosely attached to the Nb thus acting only to locally reduce the pairing amplitude but not to modify the OP symmetry.

To further verify that the ZBCPs do not reflect other possible types of impurity-induced zero-energy bound states, unrelated to the chiral molecules, we measured three control samples, and none of them showed ZBCPs. Figure 5(a) presents a typical spectrum acquired from a Nb film that has undergone the same treatment as samples containing chiral molecules, including plasma ashing, except for the final adsorption step in which it was soaked in ethanol, but *without* the chiral molecules. Evidently, no unique spectral features emerged in the spectrum, which fits well a pure s-wave simulation with $\Delta$ = 1.1 meV. Next, we turned to examine whether non-chiral organic molecules can give rise to ZBCPs. Repeating the work presented in Ref. [16], a Nb film ($T_C$ = 7.5 K) with adsorbed diSilane molecules was measured. Here too, only conventional gap structures were observed that fit only an s-wave pairing-potential with $\Delta$ = 1.2 meV (figure 5(b)), with no signature for molecular (due to the much larger HOMO-LUMO gap compare to the $\Delta$. Figure 5(c) demonstrates a spectrum measured on single crystal Nb sample ($T_C$ = 9 K), showing a gap of $\Delta$ = 1.55 meV. The spectra obtained on the single crystal sample exhibit wider and deeper gaps, with sharper coherence peaks, compared to those presented in figures 4(a) and 4(b). Therefore, the effect of disorder resulting from the specified surface treatments is only to smear the gap structure, but not to induce in-gap bound states.

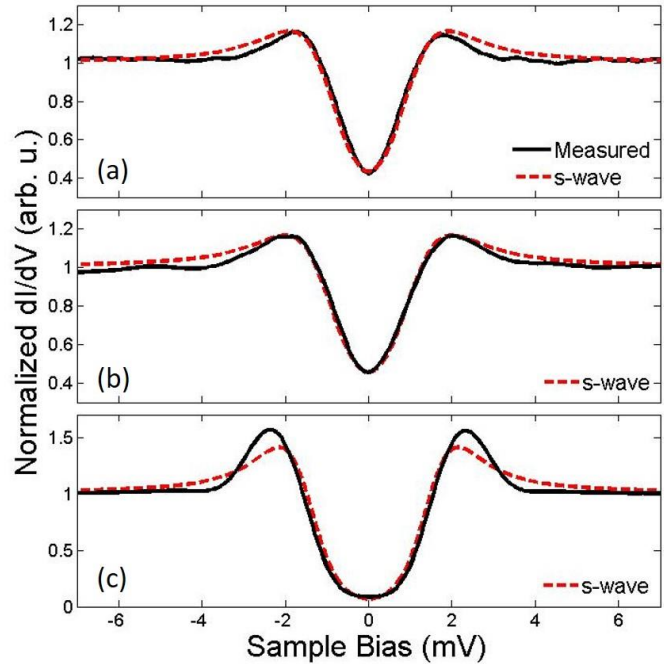

*Figure 5*: *Tunneling spectra measured on three different control samples (solid black lines) with respective fits to conventional s-wave order parameter (red dashed lines). (a) Nb film ($T_C$= 8.5 K) after 12 s oxygen plasma treatment and 24 h soaking in ethanol solution (with no molecules); Δ=1.1meV. (b) Nb film ($T_C$= 7.5 K) with non-chiral diSilane molecules adsorbed; Δ=1.2meV. (c) Bare Nb single crystal ($T_C$= 9 K); Δ=1.6 meV.*

The mechanism underlying the emergence of nonconventional OP upon chiral molecule adsorption, at least on the Nb surface, is not yet clear to us. However, in view of the previous work described above demonstrating spin-selective transport through chiral molecules [13,14] and the consequent ability to magnetize connected materials [15], it is plausible that a related spin-polarizing effect can act also on the Nb surface. Such effect may give rise to the emergence of triplet superconductivity at the surface. Support for this scenario is provided by the relative weights of the m=1 triplet-pairing p-wave contributions to the pairing-potential derived from the fits (e.g., figures 4(b,c)). These range from 60% to 75%, comparable to the spin-polarization efficiency observed in transport and optical measurements performed on the same polyalanine molecules[15]. The spectra calculated with a d-wave pairing-potential provided equally good fits to experiment, but with somewhat smaller (yet still plausible within this model) relative weights (figures 4(b,c)). While the d-wave order-parameter is commonly associated with singlet-pairing superconductivity, it can also go along with a triplet state, in the presence of odd-frequency superconductivity [35]. We do not know, however, to what extent does the odd-frequency d-wave DoS differ from its (more conventional) even-frequency counterpart. It should be noted here that proximity-induced modifications of the OP symmetry can take place also via an insulating layer (as studied here). This was shown theoretically[36] for an insulating ferromagnetic film that takes part in inducing triplet-superconductivity in both normal and singlet-superconducting regions separated by it. STS measurements on Ag/EuS/Al multilayers appear to provide evidence for this prediction[37].

Another possible mechanism is related to theoretical predictions regarding the appearance of a triplet component in a two-dimensional superconductor [38], or at a superconductor/normal-metal interface[39], lacking inversion (or mirror) symmetry in the presence of strong Rashba type spin-orbit coupling. Under these conditions, mixing of excitations from the two spin bands takes place in a manner that makes spin a non-conserved quantity. This yields an intrinsic mixture of singlet and spin-aligned triplet-pair correlations, akin to the case of heavy fermion superconductors described above. The (m=1) triplet state derived in these works is associated with p-wave orbital symmetry, consistent with our s-wave + chiral p-wave fits. While our samples do not exactly match the model systems treated in these two papers, it does resemble them. Nb has relatively strong spin-orbit coupling, which can be further enhanced by the adsorbed chiral molecules that also have a strong dipole moment. It was also shown that adsorbed chiral molecules can induce chirality in the substrate [40–42], which can further promote the emergence of a triplet chiral p-wave component.

In conclusion, our scanning tunneling spectroscopy measurements seem to demonstrate the emergence of unconventional superconductivity in the surface of Nb films upon adsorption of chiral ployanaline alpha-helix molecules. The BCS-like gap structures measured on the pristine Nb films are replaced by spectra showing ZBCPs inside gaps. Fits to simulated spectra suggest that the chiral-molecule induced pairing-potential have either triplet chiral p-wave or (singlet or triplet) d-wave symmetry, residing along with the original singlet s-wave superconductivity. While our fits cannot differentiate between the different options, it appears that theories addressing related systems, in which strong spin-orbit exists in the presence of broken inversion symmetry, favor the triplet p-wave scenario. This may be appealing to the development of chiral-based simple superconducting spintronic devices.

**Acknowledgements**

The research was supported in parts by the Leverhulme Trust, grant no. IN-2013-033; a grant from the Academia Sinica – Hebrew University Research Program (O.M. and Y.P.); and by the ISF (grant no. 1248/10). O.M. thanks support from the Harry de Jur Chair in Applied Science.